\documentclass[12pt]{article}
\usepackage[utf8]{inputenc}

\usepackage{scicite}
\usepackage{amsfonts}
\usepackage{times}
\usepackage[usenames]{color} %color
\usepackage{float}
\usepackage{graphicx}
\usepackage{amsmath,amsthm}
\usepackage{amssymb}
\usepackage{caption}
\usepackage{subcaption}
\usepackage{amsfonts}
\usepackage{amscd}
\usepackage{wrapfig}
\usepackage{boxedminipage}
\usepackage{hyperref}
\usepackage{tikz}

\theoremstyle{definition}

 \newtheorem{definition}[equation]{Definition}
 \newtheorem{remark}[equation]{Remark}
 \newtheorem{lemma}[equation]{Lemma}
 
 \newtheorem{example}[equation]{Example}
 \newtheorem{direction}{Direction}

\renewcommand{\phi}{\varphi}
\newcommand{\ZZ}{\mathbb Z}

\newenvironment{sciabstract}{%
\begin{quote} \bf}
{\end{quote}}

\newcounter{lastnote}

\title{Self-Organized Criticality and Pattern Emergence through the lens of Tropical Geometry} 
\author
{N. Kalinin,$^{1}$ A. Guzm\'{a}n-S\'{a}enz,$^{2}$ Y. Prieto, $^{3}$ M. Shkolnikov, $^{4}$ V. Kalinina,$^{5}$   E. Lupercio $^{6\ast}$\\
\\
\normalsize{$^{1}$Higher School of Economics, The Laboratory of Game Theory and Decision Making }\\
\normalsize{Saint-Petersburg, Russia}\\
\normalsize{$^{2}$Computational Genomics, IBM T.J. Watson Research Center, }\\
\normalsize{Yorktown Heights, NY, USA}\\
\normalsize{$^{3}$Institut de Mathématiques de Toulouse, Université de Toulouse III Paul Sabatier}\\ 
\normalsize{Toulouse, France}\\
\normalsize{$^{4}$Institute of Science and Technology Austria}\\
\normalsize{Klosterneuburg, Austria}\\
\normalsize{$^{5}$Institute of Cytology, Russian Academy of Science }\\
\normalsize{Saint-Petersburg, Russia}\\
\normalsize{$^{6}$CINVESTAV, Department of Mathematics }\\
\normalsize{Campus Zacatenco, CDMX, Mexico}\\
\normalsize{$^\ast$To whom correspondence should be addressed; e-mail:  elupercio@gmail.com}
}

\date{}

\begin{document} 
\baselineskip24pt
\maketitle 
\begin{sciabstract}
%In this paper, w
Tropical Geometry, an established field in pure mathematics, is a place where String Theory, Mirror Symmetry, Computational Algebra, Auction Theory, etc, meet and influence each other. In this paper, we report on our discovery of a tropical model with self-organized criticality (SOC) behavior. Our model is continuous, in contrast to all known models of SOC, and is a certain scaling limit of the sandpile model, the first and archetypical model of SOC. We describe how our model is related to pattern formation and proportional growth phenomena, and discuss the dichotomy between continuous and discrete models in several contexts. Our aim in this context is to present an idealized tropical toy-model (cf. Turing reaction-diffusion model), requiring further investigation.

%In this paper, we put forward a new mathematical paradigm in the study of self-organized critical phenomena in nature: Tropical Algebraic Geometry. We describe our discovery of a continuous Tropical dynamical system that exhibits self-organized criticality using data generated by two super-computer clusters. We also explain how Tropical Geometry and this model in particular, are useful in the study of proportional growth phenomena.% in the classical sandpile model.   
\end{sciabstract}

\section*{Self-Organized Criticality}
The statistics concerning earthquakes in a particular extended region of the Earth during a given period of time obey a power law known as the Gutenberg-Richter law \cite{newman2005power}: the logarithm of the energy of an earthquake is a linear function of the logarithm of the frequency of earthquakes of such energy.  An earthquake is an example of a dynamical system presenting both temporal and spatial degrees of freedom.  The following definition generalizes these properties. 

\begin{definition} A dynamical system is a \emph{Self-Organized Critical} (SOC) system if it is slowly driven by an external force, exhibits avalanches, and has power law correlated interactions (cf. \cite{watkins201625}, section 7).
\end{definition}

By an avalanche, we mean a sudden recurrent modification of the internal energy of the system. In SOC systems, avalanches display scale invariance over energy and over time \cite{watkins201625}\footnote{P.W. Anderson describes SOC as having ``paradigmatic value" characterizing ``the next stage of Physics". He writes: ``In the 21$^\textit{st}$ century, one revolution which can take place is the construction of generalization which jumps and jumbles the hierarchies or generalizations which allow scale-free or scale transcending phenomena. The paradigm for the first is broken symmetry; for the second, self-organized criticality"  \cite{anderson2011more}.}.

 Many ordinary critical phenomena near continuous phase transition (which requires fine tuning) display non-trivial power law correlations with a cut-off associated to the cluster size \cite{aharony2003introduction}. For example, the Ising model shows power law correlations only for specific parameters, while a self-organized critical system, behaving statistically in an analogous manner, achieves the critical state merely by means of a small external force, without fine-tuning.

\section*{The Sandpile Model}

The concept of SOC was introduced in the seminal papers of Per Bak, Chao Tang, and Kurt Wiesenfeld \cite{BTW,bak1988self} where they put forward the archetypical example of a SOC system: the Sandpile Model.
It is a highly idealized cellular automaton designed to display spatio-temporal scale invariance. Unlike other toy models in physics, e.g. the Ising model, a sandpile automaton doesn't attempt to apprehend the actual interactions of a physical system exhibiting SOC (such as a real sandpile). 
The sandpile cellular automaton is rather a mathematical model \footnote{A holistic model in the sense of  \cite{downey2012think}.} that reproduces the statistical behavior of very diverse dynamical systems: it is descriptive at the level of global emergent phenomena without necessarily corresponding to such systems at the local reductionist level.

\begin{figure}[h]
  \begin{center}
  \begin{tikzpicture}
   \begin{scope}[scale = 0.8]
   \draw [very thick, dashed] (1,1) grid (4,4);
   \draw [red] (3,3)node{$\bullet$};
   \draw (3,3) node[above right]{4};
   \draw (2,3) node[above right]{3};
   \draw (3,2) node[above right]{3};
   \draw (4,3) node[above right]{3};
   \draw (3,4) node[above right]{0};
   \draw (2,2) node[above right]{1};
   \draw (5.7,2.7) node {$\text{toppling}$};
   \draw[->, very thick] (5,2)--(6.5,2);
   \end{scope} 
  \begin{scope}[xshift=140, scale = 0.8]
   \draw [very thick, dashed] (1,1) grid (4,4);
   \draw [red] (2,3)node{$\bullet$};
   \draw (3,3) node[above right]{0};
   \draw (2,3) node[above right]{4};
   \draw (3,2) node[above right]{4};
   \draw (4,3) node[above right]{4};
   \draw (3,4) node[above right]{1};
   \draw (2,2) node[above right]{1};
   \draw [red] (3,2)node{$\bullet$};
   \draw [red] (4,3)node{$\bullet$};
   \end{scope}

  \end{tikzpicture}
  \end{center}
\caption{Numbers represent the number of sand grains in the vertices of the grid, and a toppling is performed. Red points are unstable vertices.}
\label{pic_sand}
\end{figure}
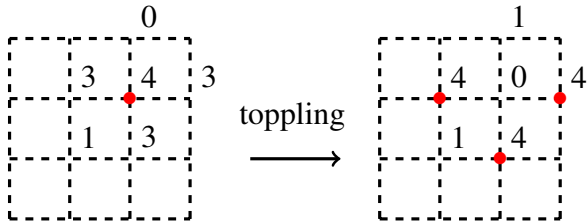

Imagine a large domain on a checkered plane; each vertex of this grid is determined by two integer coordinates $(i,j)$. We will write $\mathbb{Z}^2$ to denote the integral square lattice formed by all pairs of integers $(i,j)$ in the Euclidean plane $\mathbb{R}^2$. Our dynamical system will evolve inside $\Omega$, which is the intersection of $\mathbb{Z}^2$ with a large convex figure in the plane. Throughout this paper, we use the word \emph{graph} to indicate a collection of line segments (called edges) that connect points (called vertices) to each other.  

\begin{definition} A sandpile model consists of a grid inside a convex domain $\Omega$ on which we place grains of sand at each vertex; the number of grains on the vertex $(i,j)$ is denoted by $\varphi(i,j)$ (see Figure~\ref{pic_sand}).  Formally, a {\it state} is an integer-valued function $\varphi:\Omega\to\mathbb Z_{\geq 0}$.    We call  a vertex $(i,j)$ \emph{unstable} whenever there are four or more grains of sand at $(i,j)$, i.e., whenever $\varphi(i,j)\geq 4$. The evolution rule is as follows:  any unstable vertex $(i,j)$ topples spontaneously by sending one grain of sand to each of its four neighbors $(i, j + 1),(i,j-1),(i-1,j),(i+1,j)$. The sand that falls outside $\Omega$ disappears from the system.  Stable vertices cannot be toppled. Given an initial state $\varphi$, we will denote by $\varphi^\circ$ the stable state reached after all possible topplings have been performed. It is a remarkable and well-known fact that the final state $\phi^\circ$ does not depend on the order of topplings. The final state $\phi^\circ$ is called the relaxation of the initial state $\phi$.
\end{definition}

Bak and his collaborators proposed the following experiment: take any stable state $\phi_0$ and perturb it at random by adding a grain of sand at a random location. Denote the relaxation of the perturbed state by $\phi_1$, and repeat this procedure.  Thus a sequence of randomly chosen vertices $(i_k,j_k)$ gives rise to a sequence of stable states by the rule  $\phi_{k+1}= (\phi_k + \delta_{i_kj_k})^\circ$. 

The relaxation process $(\phi_k + \delta_{i_kj_k})\mapsto \phi_{k+1}$ is called an \emph{avalanche}\footnote{We can think of an avalanche as an earthquake.}; its size is the number of vertices that topple during the relaxation. Given a long enough sequence of uniformly chosen vertices $(i_k,j_k)$, we can compute the distribution for the sizes of the corresponding avalanches. Let $N(s)$ be the number of avalanches of size $s$; then the main experimental observation of \cite{BTW} is that:

$$
\log N(s)= \tau \log s + c.
$$ 
In other words, the sizes of avalanches satisfy a power law. In Figure~\ref{fig:SandpilePowerLaw}, we have reproduced this result with $\tau \sim -1.2$. This has only very recently been given a rigorous mathematical proof using a deep analysis of random trees on the two-dimensional integral lattice $\mathbb{Z}^2$ \cite{bhupatiraju2016inequalities}.

\begin{definition}
A recurrent state is a stable state appearing infinitely often, with probability one, in the above dynamical evolution of the sandpile. 
\end{definition}

Surprisingly, the recurrent states are exactly those which can be obtained as a relaxation of a state $\phi\geq 3$ (pointwise).
The set of recurrent states forms an Abelian group \cite{Dhar} and its identity exhibits a fractal structure in the scaling limit (Fig.~\ref{fig:PhaseTransition}); unfortunately, this fact has resisted a rigorous explanation so far.

The main point of this paper is to exhibit fully analogous phenomena in a continuous system (which is not a cellular automaton) within the field of tropical geometry. 

An advantage of the tropical model is that, while it has self-organized critical behavior, just as the classical model does, its states look much less chaotic; thus we say that the tropical model has no combinatorial explosion.

\section*{Mathematical Modeling for Proportional Growth and Pattern Formation}

The dichotomy between continuous mathematical models and discrete cellular automata has an important example in developmental biology.

A basic continuous model of pattern formation was offered by Alan Turing in 1952 \cite{turing1952chemical}. He suggested that two or more homogeneously distributed chemical substances, termed morphogens, with different diffusing rates and chemical activity, can self-organize into spatial patterns of different concentrations. This theory was confirmed decades later and can be applied, for example, to modeling the patterns of fish and lizard skin \cite{kondo1995reaction},\cite{dhillon2017bifurcation}, or of seashells' pigmentation \cite{fowler1992modeling}.

On the discrete modeling side, the most famous model is the Game of Life developed by Conway in 1970 \cite{gardner1970mathematical}. A state of this two-dimensional system consists of "live" and "dead" cells that function according to simple rules. Any live cell dies if there are fewer than two or more than three live neighbors. Any dead cell becomes alive if there are three live neighbors. A very careful control of the initial state produces "oscillators" and "spaceships" that certainly look fascinating but seem not to be related to realistic biological models. Nevertheless, a strong philosophical conclusion of the Game of Life is that extremely simple rules can produce behavior of arbitrary complexity. A more realistic cellular automaton has recently been derived from  the continuous reaction-diffusion model of Turing. In \cite{manukyan2017living},  the transformation of the juvenile white ocelli skin patterns of the lizard {\it Timon lepidus} into a green and black labyrinth was observed. In this study, the authors presented the skin squamae of lizard as a hexagonal grid, where the color of each individual cell depended on the color states of neighboring positions. This cell automaton could successfully generate patterns coinciding with those on the skin of adult lizards.

\begin{figure}[H]
	\centering
	\includegraphics[width=1.0\linewidth]{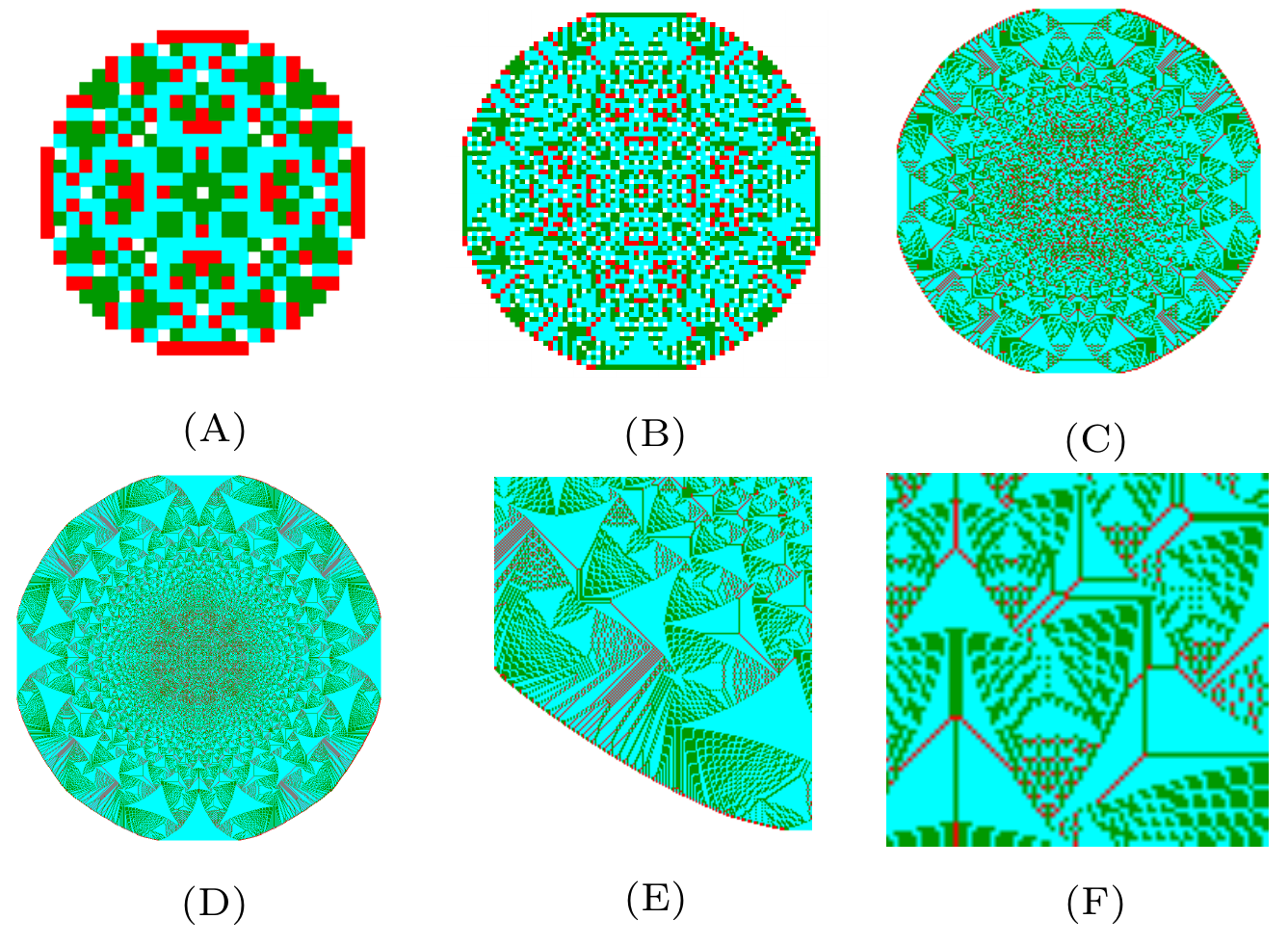}
		
	\caption{ In (A), (B), (C) and (D), a very large number $N$ of grains of sand is placed at the origin of the everywhere empty integral lattice, the final relaxed state shows fractal behavior. Here, as we advance from (A) to (D), we see successive sandpiles for $N=10^3$ (A), $10^4$ (B), $10^5$ (C), and $10^6$ (D), rescaled by factors of $\sqrt{N}$. In (E), we zoom in on a small region of (D) to show its intricate fractal structure, and, finally, in (F), we further zoom in on a small portion of (E). We can see proportional growth occurring in the patterns as the fractal limit appears. The balanced graphs inside the roughly triangular regions of (F) are tropical curves.}
	\label{fig:scaleinvariance}
\end{figure}

Pattern formation is related to an  important problem in developmental biology: how to explain proportional growth.  It is not clear  why  different parts of animal or human bodies grow  at approximately the same rate from birth to adulthood. Sandpile simulations provide a qualitative toy model as follows.

\begin{example}
\label{ex_many}
Early on, it was observed experimentally that sandpiles have fractal structure in their spatial degrees of freedom (see  Figure~\ref{fig:scaleinvariance}).  
\end{example}
This example exhibits the phenomenon of proportional growth and scale invariance: If we rescale tiles to have  area $\frac{1}{N}$ and let $N$ go to infinity, then the picture converges in the weak-$\star$ sense. (See \cite{PS} and references therein.) Recently, the patches and a linear pattern in this fractal picture were explained in \cite{LPS,us_solitons,levine2013apollonian} using discrete superharmonic functions and Apollonian circle packing.

Dhar and Sadhu \cite{dhar2013sandpile} established certain two-dimensional sandpile models where the size of newly formed patterns depends on the number of sand grains added on the top, but the number and shape of the patterns remain the same. Strikingly, they also proposed a three-dimensional model which also forms proportionally growing patterns; these patterns look like the  head, thorax, and abdomen of real larva. 

The perspective of our paper suggests that continuous tropical geometry should have consequences in the study of proportional growth modeling. We conjecture that tropical functions should appear as gradients of growth.(Compare Figure 3 in ~\cite{rastelli2002surface} and our Figure~\ref{figpush}; see also Section 2.5 in \cite{vollmer2017growth}). 

\section*{Tropical Geometry}

Tropicalization can be thought of as the study of geometric and algebraic objects on the log-log scale\footnote{Drawing algebraic varieties on log-log scale was successfully used by O. Ya. Viro, in his study of Hilbert's 17th problem, to construct real algebraic curves with prescribed topology, for a more recent account of this story read \cite{viro2006patchworking}.} in the limit when the base of the logarithm is very large\footnote{This limit has been discovered several times in physics and mathematics and was named Maslov dequantization in mathematical physics before it was called tropicalization in algebraic geometry \cite{kolokoltsov1997idempotent}.}. Let us start by considering the most basic mathematical operations: addition and multiplication. With the logarithmic change of coordinates, and with the base of the logarithm becoming infinitely large, multiplication becomes addition and addition becomes taking the maximum.  Namely, define

$$x+_t y := \log_{t}(t^x+t^y)$$
$$x \times _t y := \log_t(t^xt^y)$$
for $x, \ y$ positive and then, taking the limit as $t$ tends to $+ \infty$,  set
$${\mathrm{Trop}}(x+y):= \lim\limits_{t \rightarrow + \infty} x +_t y = \max(x,y),$$
$${\mathrm{Trop}}(x \times y):= \lim\limits_{t \rightarrow + \infty} x \times_t y = x+y.$$
Tropical numbers are, thus, defined as the set of ordinary real numbers (together with $\{- \infty \}$, i.e. $\mathbb{T} := \mathbb{R} \cup \{ - \infty \}$) and with tropical addition $\max(x,y)$ and multiplication $x+y$. The tropical numbers form a semi-ring:  all the usual rules of algebra hold in $\mathbb{T}$ with the exception of additive inverses (the neutral element for addition is $-\infty$; additive inverses are missing).

This mathematical structure simplifies some computations, and several optimization problems are solved efficiently in this way; this is a growing area of applied research in the current decade (it has been used in auctions for the Bank of England \cite{economy2}, for biochemical networks \cite{radulescu2012reduction}, and for Dutch railroads, among other things \cite{heidergott2014max}).

Thermodynamics suggests that the tropical limit should be understood as the low temperature limit $T\to 0$ (or equivalently, as the limit when the Boltzmann constant $k$ vanishes, $k\to 0$) of the classical algebraic and geometric operations. In this interpretation of the limit, the relevant change of variables is $t=e^{-\frac{1}{kT}}$, where $t$ is the tropical parameter (the base of the logarithm) and $T$ is the temperature \cite{kenyon2006dimers,kapranov2011thermodynamics,itenberg2012Geometry}. More relevant to realistic physical systems, tropical functions and tropical algebra appear naturally in statistical physics as the formal limit as $k\to 0$, and this can be used to analyze frustrated systems like spin ice and spin glasses. This is directly related to the appearance of the tropical degeneration in dimer models \cite{kenyon2006dimers,cimasoni2007dimers}.

\begin{figure}[H]
	\centering
	\includegraphics[width=0.8\linewidth]{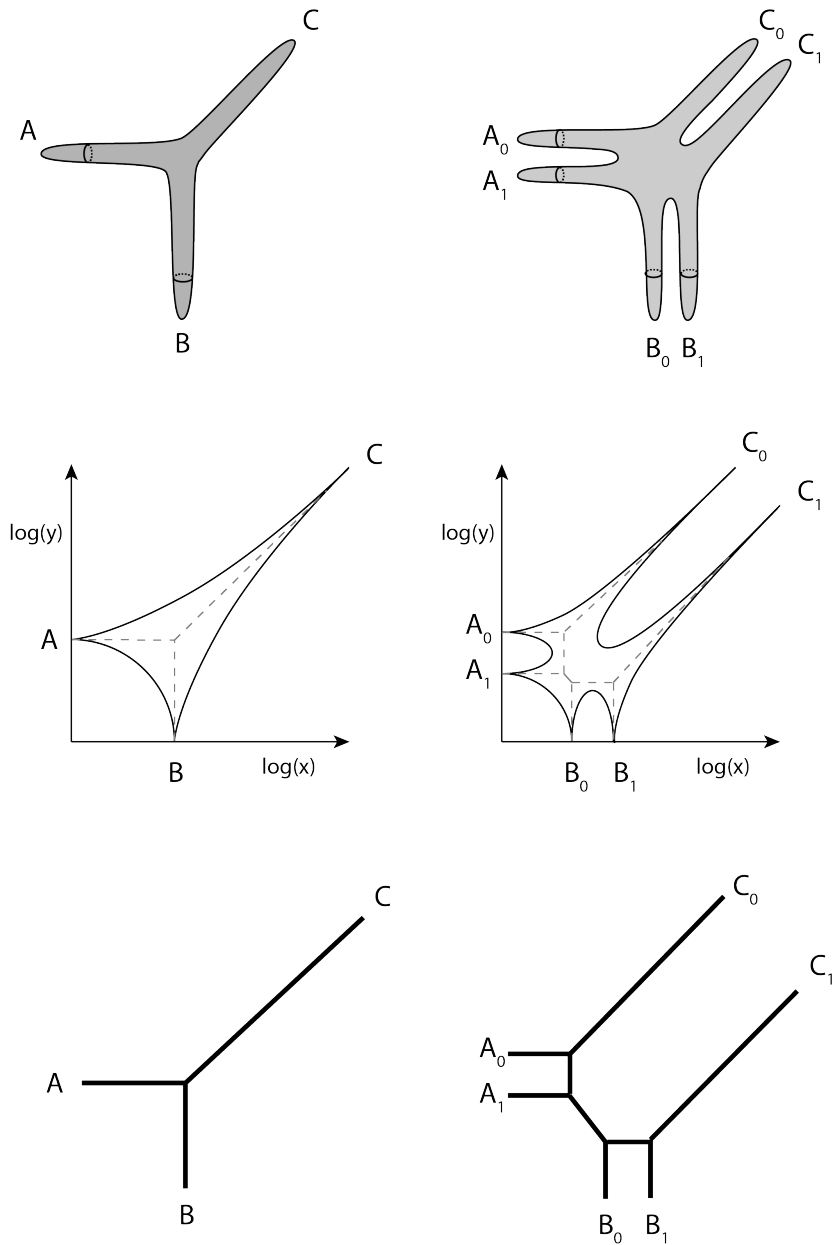}	
	\caption{Each column represents a tropical degeneration (from top to bottom). The complex picture is at the top; in the middle, we depict the amoeba on log-log scale and, at the bottom, we picture the tropical curve which can be seen as the spine of the corresponding amoeba. In the first example, we have a degree-one curve with three points $A,B,C$ going off to infinity (the intersections $A,B$ of a line with the coordinate axes go to $-\infty$ under the logarithm, and $C$, the intersection at infinity, goes to $(+\infty,+\infty)$). In the second example, a quadric degenerates sending off six  points off to infinity. A quadric intersects each of coordinate lines and the line at infinity in two points, therefore, six (three times two) points go off to infinity. From top to bottom $t\to\infty$}
	\label{fig:Tropicaldegeneration}
\end{figure}

\begin{definition}
	Let $A\subset\ZZ^2$ be any finite set. A tropical polynomial in two variables is a function
	\begin{equation}
	\label{eq_1}
	\mathrm{Trop}(F(x,y))= \max \limits_{(i,j)\in \mathcal{A}}(a_{ij}+ix+jy).
	\end{equation}

\end{definition}

A tropical polynomial should be thought of the tropicalization of a classical polynomial $F(x,y)=\sum\limits_{(i,j)\in A} a_{ij}x^iy^j$ obtained by replacing all  of the summations and multiplications by their tropical counterparts, i.e. $\max$ and addition, respectively.

For an excellent introduction to tropical geometry, we refer the reader to \cite{maclagan2015introduction}.

\section*{Tropical Limit in Algebraic Geometry}
Complex algebraic geometry is the field of mathematics  that studies geometric objects inside complex Euclidean spaces $\mathbb{C} ^n$
that arise as zero sets of finite families of polynomials. A simple example of such an object is an algebraic curve $\mathcal{C}$ given by a single polynomial equation in two variables
$$\mathcal{C}: F(x,y)=0, \ \ (x,y) \in \mathbb{C}^2, F(x,y)= \sum \limits_{(i,j)\in \mathcal{A}} a_{ij}x^iy^j.$$

Here $\mathcal{A}$ is a finite set of pairs of positive integers. The degree of $F$ is the maximal total power $i+j$ of all the monomials $x^i y^j$ appearing in $F$, namely $d=\max \limits_{(i,j)\in \mathcal{A}}(i+j)$. 
The curve $\mathcal{C}$ is a two-dimensional object, despite its name (for two real dimensions is the same as one complex dimension).

A non-singular complex curve $\mathcal{C}= \{F(x,y)=0 \}$ of degree $d=\textrm{deg}(F)$ is a Riemann surface of genus $g=\frac{1}{2}(d-1)(d-2)$. Therefore, the usual lines ($d=1$) and quadrics ($d=2$) are topological $2$-spheres, while cubics ($d=3$, also called elliptic curves) are tori.

The geometric counterpart of the tropicalization is as follows. Given a complex algebraic curve $\mathcal{C}_t$ defined by a polynomial $$F_t(x,y)=\sum \limits_{(i,j)\in \mathcal{A}}\gamma_{ij}t^{a_{ij}}x^iy^j=0, |\gamma_{ij}|=1,$$ we call the {\it amoeba} $A_t$ the image of $\mathcal{C}_t$ under the map $\log_t(x,y)=(\log_t | x | , \log_t | y |)$, $A_t:=\log_t(\mathcal{C}_t)$. The limit of the amoebas $A_t$ as $t \rightarrow +\infty$ is called ${\mathrm{Trop}}(\mathcal{C})$, the tropicalization of $\mathcal{C}_t$.

The limit ${\mathrm{Trop}}(\mathcal{C}) $
can be described entirely in terms of the tropical polynomial ${\mathrm{Trop}}(F)$ (eq.~\ref{eq_1}). 
This fact can be proved by noticing that on the linearity regions of ${\mathrm{Trop}}(F(x,y))$, one monomial in $F_t$ dominates all the others and, therefore, $F_t$ cannot be zero; and, consequently, we conclude that the limit ${\mathrm{Trop}}(\mathcal{C} )$ is precisely the set of points $(x,y)$ in the plane where the (3-dimensional) graph of the function $${\mathrm{Trop}}(F(x,y)) = \max \limits_{(i,j)\in \mathcal{A}}(a_{ij}+ix+jy)$$
is not smooth. This set of points is known as the corner locus of ${\mathrm{Trop}}(F(x,y))$. For this reason, we define the tropical plane curve as the corner locus of a tropical polynomial. As depicted in Figure~\ref{fig:Tropicaldegeneration}, the tropical degeneration ${\mathrm{Trop}}(\mathcal{C})$ of a complex curve $\mathcal{C}$ is essentially the graph of the curve depicted in  a $\log_t$-$\log_t$ scale for $t$ very large.

It is not very hard to verify that every tropical curve ${\mathrm{Trop}}(\mathcal{C})$ in the plane is a finite balanced graph, all of whose edges have rational slopes, such that at every vertex, the (weighted) directions of the primitive vectors in every direction cancel out (this is called the balancing condition and is akin to Kirchhoff's law for electrical circuits). Conversely, every such balanced graph on the plane appears as a tropical algebraic curve ${\mathrm{Trop}}(\mathcal{C})$. Here, it may be a good moment to point out that one can observe small tropical curves in Figure \ref{fig:scaleinvariance}, where the final state of a sandpile is depicted; this already points towards a relationship between SOC and tropical geometry.

\section*{String Theory, Mirror Symmetry}

Tropical geometry is intimately related to the interaction between algebraic geometry and string theory that occurs in the mirror symmetry program. Given a tropical 2-dimensional surface $B$, we can use the tropical structure to produce a pair $(X_B,\tilde{X}_B)$ of mirror manifolds. This motivated M. Kontsevich to predict that counting tropical curves on $B$ could be used for the calculation of Gromov-Witten invariants (which count holomorphic complex curves) in the $A$-side of mirror symmetry. The first example of such a calculation done from a rigorous mathematical perspective was accomplished by Mikhalkin \cite{mikh1}. This perspective has been expanded by Gross and Siebert  in their program to understand mirror symmetry from a mathematical viewpoint using tropical geometry \cite{MR2722115}. 

Let us also mention that the dichotomy between continuous and discrete models in our paper (already appearing in the biological models) has an important analogue in string theory: Iqbat et al. have argued that, when we probe space-time beyond the scale $\alpha'$ and below Planck's scale, the resulting fluctuations of space-time can be computed with a classical cellular automaton (a melting crystal) representing quantum gravitational foam \cite{MR2425292}. Their theory is a three-tier system whose levels are classical geometry (K\"ahler gravity), tropical geometry (toric manifolds) and cellular automata (discrete melting crystals). The theory that we propose in this paper is also a three-tier system whose levels are classical complex algebraic geometry, tropical geometry (analytic tropical curves) and cellular automata (sandpiles). This seems not to be a coincidence and suggests deep connections between our model for SOC and their model for quantum gravitational foam.

\section*{Tropical Curves in Sandpiles} 	   

To understand the appearance of tropical geometry in sandpiles, consider the \emph{toppling function} $H(i,j)$ defined as follows: Given an initial state $\phi$ and its relaxation $\varphi^\circ$, the value of $H(i,j)$ equals the number of times that there was a toppling at the vertex $(i,j)$ in the process of 
taking $\varphi$ to $\varphi^\circ$. 

The discrete Laplacian of $H$ is defined by the net flow of sand, $\Delta H (i,j) :=$
$$ H(i-1,j) + H(i+1,j) + H(i,j-1) + H(i,j+1) - 4 H(i,j).$$

The toppling function is clearly non-negative on $\Omega$ and vanishes at the boundary. The function $\Delta H$ completely determines the final state $\varphi^\circ$ by the formula:
\begin{equation}
\label{eq_topp}
\varphi^\circ(i,j) = \varphi (i,j) + \Delta H(i,j).
\end{equation}

\begin{figure}[H]
\setlength{\fboxsep}{0pt}
	\centering
	\includegraphics[width=1.0\linewidth]{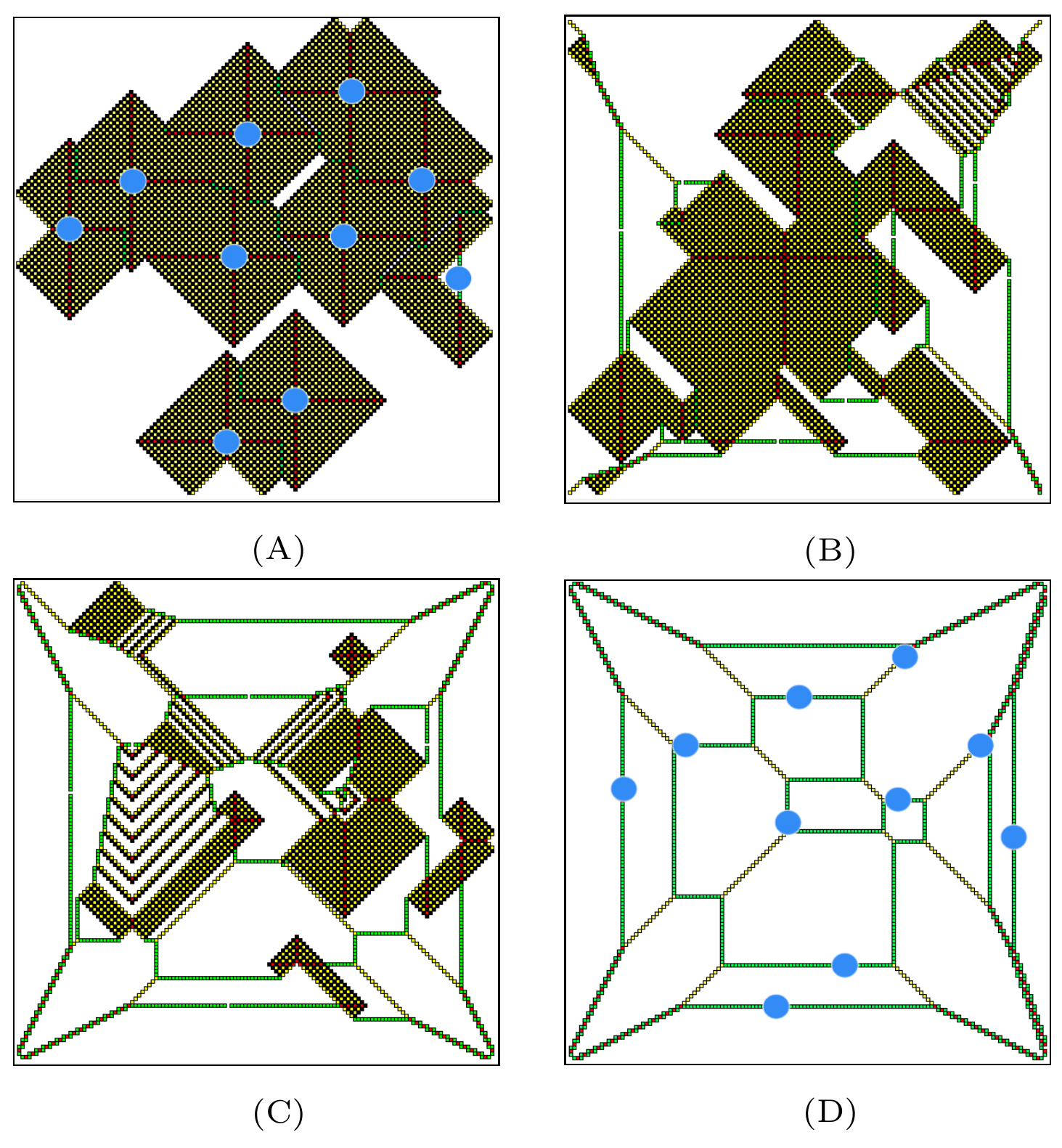}
	\caption{The evolution of $\langle 3 \rangle + \delta_{P}$. Sand falling outside the border disappears. Time progresses in the sequence (A), (B), (C), and finally (D). Before (A), we add grains of sand to several points of the constant state $\langle 3\rangle$ (we see their positions as blue disks given by $\delta_P$). Avalanches ensue. At time (A), the avalanches have barely started. At the end, at time (D), we get a tropical analytic curve on the square $\Omega$. White represents the region with 3 grains of sand while green represent 2, yellow represents 1, and red represents the zero region. We can think of the blue disks $\delta_P$ as the genotype of the system, of the state $\langle 3 \rangle$ as the nutrient environment, and of the thin graph given by the tropical function in (D) as the phenotype of the system.}
	\label{fig:Tropicalsandpileevolution}
\end{figure}

It can be shown by induction that the toppling function $H$ satisfies the \emph{Least Action Principle}: if $\varphi(i,j) + \Delta F(i,j) \leq 3$ is stable, then $F(i,j)\geq H(i,j)$. Ostojic \cite{ostojic2003patterns} noticed that $H(i,j)$ is a piecewise quadratic function in the context of Example~\ref{ex_many}.

Consider a state $\varphi$ which consists of 3 grains of sand at every vertex, except at a finite family of points $$P=\{p_1=(i_1,j_1),\ldots,p_r=(i_r,j_r)\}$$ where we have 4 grains of sand:
\begin{equation}
\label{eq_phi}
\varphi:=\langle 3 \rangle + \delta_{p_1} + \cdots + \delta_{p_r} = \langle 3 \rangle + \delta_{P} .
\end{equation}

The state $\varphi^\circ$ and the evolution of the relaxation can be described by means of tropical geometry (the final picture (D) of Figure~\ref{fig:Tropicalsandpileevolution} is a tropical curve). This was discovered by Caracciolo et al. \cite{caracciolo2010conservation} while 
a rigorous mathematical theory to prove this fact has been given by Kalinin et al. \cite{us}, which we review  presently. It is a remarkable fact that, in this case, the toppling function $H(i,j)$ is piecewise linear (after passing to the scaling limit).

To prove this, one considers the family $\mathcal{F}_P$ of functions on $\Omega$ that are: (1) piecewise linear with integral slopes, (2) non-negative over $\Omega$ and zero at its boundary, (3) concave, and (4) not smooth at every point $p_i$ of $P$.
Let $F_P$ be the pointwise minimum of functions in $\mathcal{F}_P$. 
Then  $F_P \geq H$ by the Least Action Principle (since $\Delta F_P\leq 0, \Delta F_P(p_i)<0$). 

\begin{lemma}
In the scaling limit $H=F_P$.
\end{lemma}
{\bf A sketch of a proof.} We introduce the wave operators $W_p$ \cite{ivashkevich1994waves,ktitarev2000scaling}  at the cellular automaton level  and the corresponding tropical wave operators $G_p$. Given a fixed vertex $p=(i_0,j_0)$, we define the wave operator $W_p$ acting on states $\varphi$ of the sandpile as:
$$ W_p(\varphi):=(T_p(\varphi+\delta_p)-\delta_p)^{\circ},$$
where $T_p$ is the operator that topples the state $\varphi+\delta_p$ at $p$ once, if it's possible to topple $p$ at all. In a computer simulation, the application of this operator looks like a wave of topplings spreading from $p$, while each vertex topples at most once.

The first important property of $W_p$ is that, for the initial state $\varphi:=\langle 3 \rangle + \delta_{P}$, we can achieve the final state $\varphi^\circ$ by successive applications of the operator $W_{p_1}\circ\cdots\circ W_{p_r}$ a large but finite number of times (in spite of the notation):
$$\varphi^\circ = (W_{p_1}\cdots W_{p_r})^\infty \varphi+\delta_P.$$
This process decomposes the total relaxation $\varphi \mapsto \varphi^\circ$ into layers of controlled avalanching.

The second important property of the wave operator $W_p$ is that its action on a state $\varphi = \langle 3 \rangle + \Delta f$ has an interpretation in terms of tropical geometry. To wit, whenever $f$ is a piecewise linear function with integral slopes that, in a neighborhood of $p$, is expressed as
$a_{i_0j_0} + i_0x + j_0 y$, we have $$W_p(\langle 3 \rangle +\Delta f) = \langle 3 \rangle + \Delta W(f),$$ where $W(f)$ has the same coefficients $a_{ij}$ as $f$ except one, namely $a_{i_0j_0}'=a_{i_0j_0}+1$. This is to account for the fact that the support of the wave is exactly the face where $a_{i_0j_0} + i_0x + j_0 y$ is the leading part of $f$.  

\begin{figure}[H]
	\centering
	\includegraphics[width=1.0\linewidth]{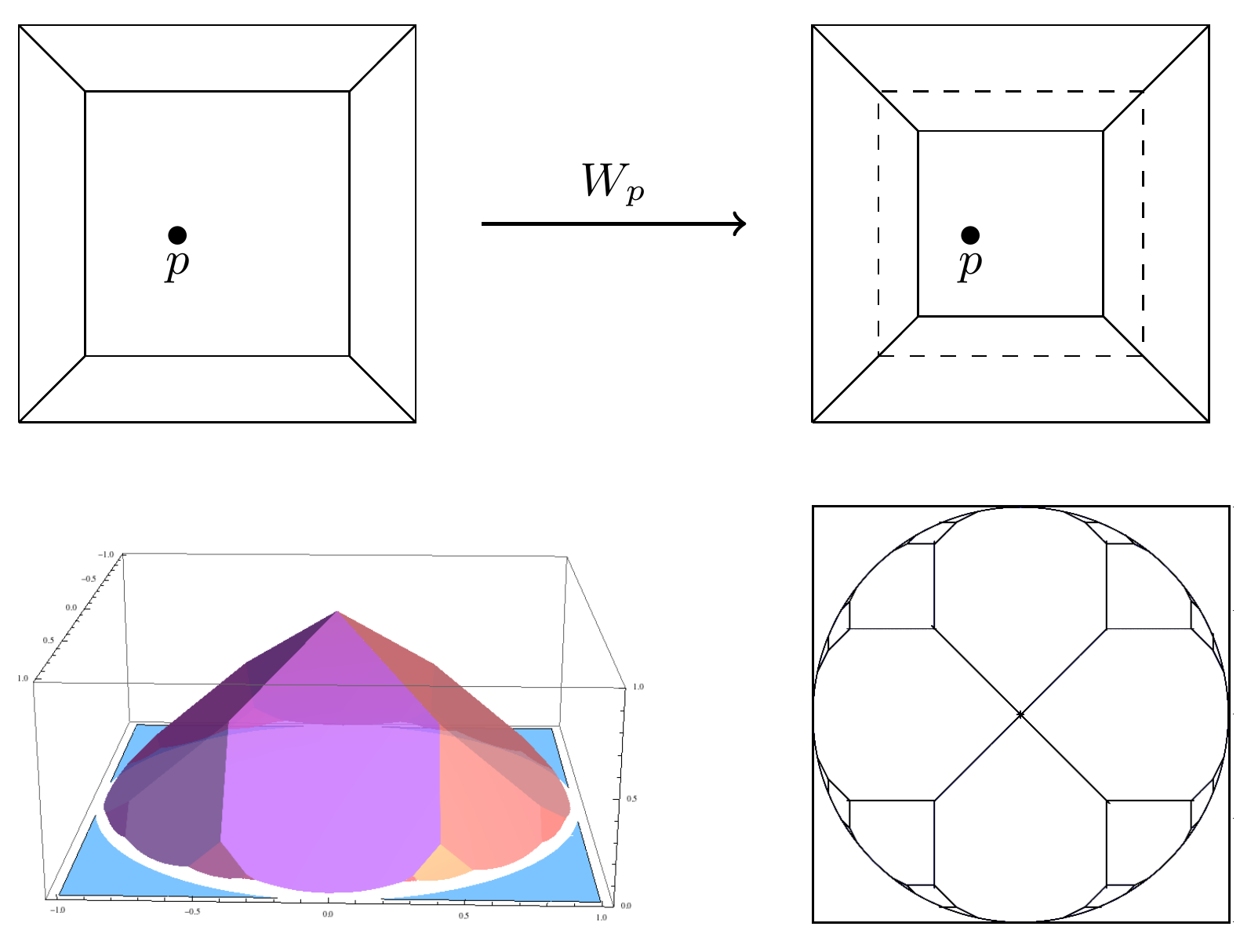}

\caption{Top: The action of the wave operator $W_p$ on a tropical curve. The tropical curve steps closer to $p$ by an integral step. Thus $W_p$ shrinks the face that $p$ belongs to; the combinatorial morphology of the face that $p$ belongs to, can actually change. Bottom: The function $G_p{\mathbf 0}$, where $p$ is the center of the circle, and its associated omega-tropical curve are shown.
}
\label{figpush}
\end{figure}

\begin{figure}[H]
	\centering
	\includegraphics[width=1.0\linewidth]{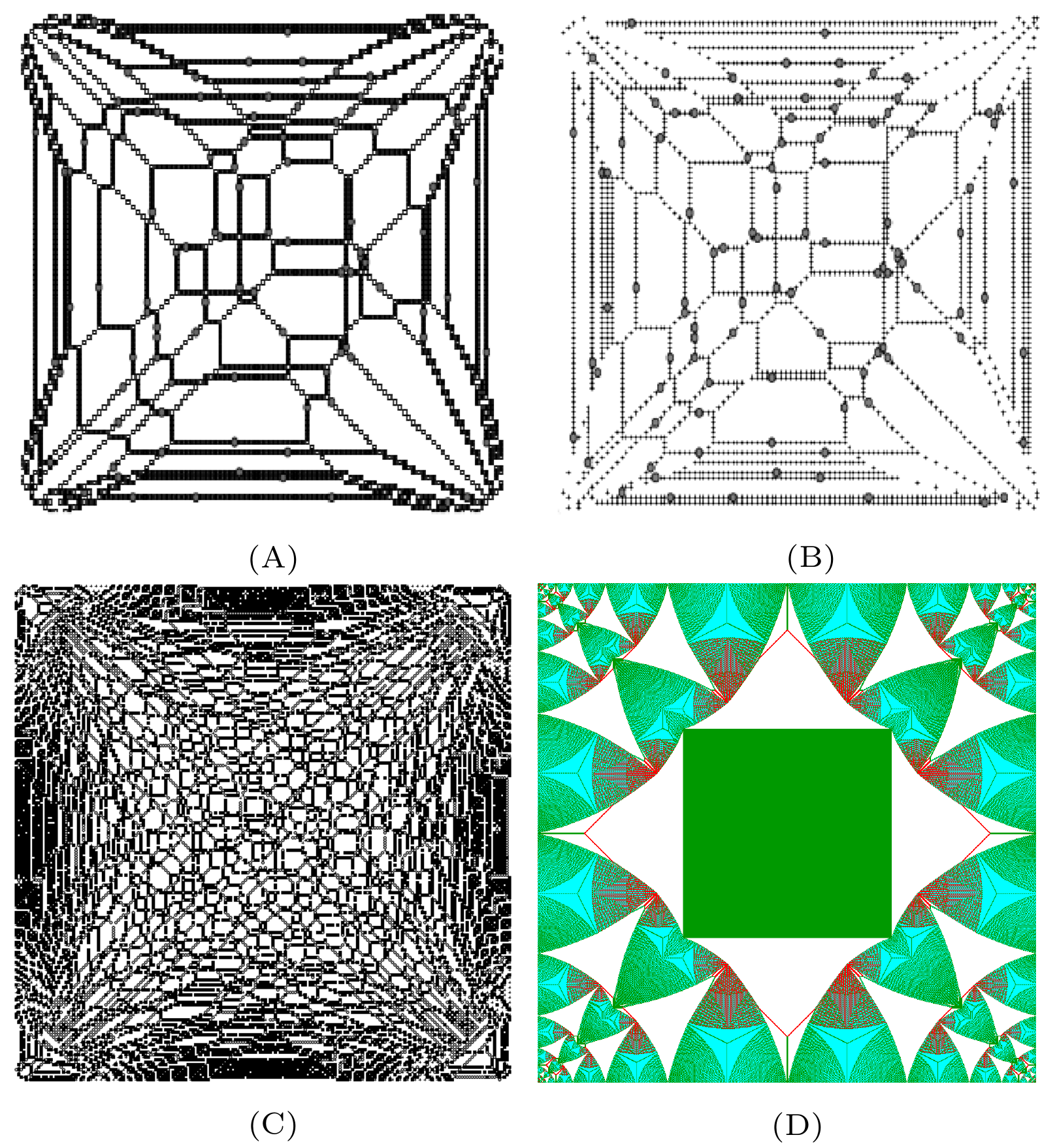}
	\caption{The first two pictures show the comparison between the classical (A) and tropical (B) sandpiles for $|P|=100$ generic points on the square. In (C), the square $\Omega$ has side $N=1000$; a large number ($|P|=40000$) of grains has been added, showing the spatial SOC behavior on the tropical model compared to the identity (D) of the sandpile group on the square of side $N=1000$. In the central square region on (C) (corresponding to the solid block of the otherwise fractal unit), we have a random tropical
curve with edges on the directions $(1, 0), (0, 1)$, and $(\pm 1, 1)$, which is given by a small
perturbation of the coefficients of the tropical polynomial defining the usual square grid.
}
\label{fig:PhaseTransition}
\end{figure}

We will write $G_p := W_p^\infty$ to denote the operator that applies $W_p$ to $\langle 3 \rangle + \Delta f$ until $p$ lies in the corner locus of $f$.We repeat that it has an elegant interpretation in terms of tropical geometry: $G_p$ increases the coefficient $a_{i_0j_0}$ corresponding to a neighborhood of $p$, lifting the plane lying above $p$ in the graph of $f$ by integral steps until $p$ belongs to the corner locus of $G_pf$. Thus $G_p$ has the effect of pushing the tropical curve  towards $p$ (Figure~\ref{figpush}) until it contains $p$.

From the properties of the wave operators, it follows immediately that:
$$F_P= \left(G_{p_1}\cdots G_{p_r}\right)^\infty {\mathbf 0},$$
where ${\mathbf 0}$ is the function which is identically zero on $\Omega$.  All intermediate functions $\left(G_{p_1}\cdots G_{p_r}\right)^k {\mathbf 0}$ are less than $H$ since they represent partial relaxations, but their limit belongs to $\mathcal{F}_P$, and this, in turn, implies that $H=F_P$. 

{\bf Conclusion.} We have shown that the toppling function $H$ for Eq.~\ref{eq_phi} is piecewise linear. Thus, applying Eq.~\ref{eq_topp}, we obtain that $\phi^\circ$ is equal to $3$ everywhere but the locus where $\Delta H\ne 0$, i.e. its corner locus, namely, an $\Omega$-tropical curve (Figure~\ref{fig:Tropicalsandpileevolution}).

\begin{definition}\cite{us_series}
An $\Omega$-tropical series is a piecewise linear function on $\Omega$ given by:
$$F(x,y) = \min_{(i,j)\in\mathcal{A}} (a_{ij} +ix + jy),$$
where the set $\mathcal{A}$ is not necessarily finite and $F|_{\partial\Omega}=0$. An $\Omega$-tropical curve is the set where $F$ is not smooth. Each $\Omega$-tropical curve is a locally finite graph satisfying the balancing condition.

\end{definition}

\begin{remark}
Tropical curves consist of edges, such that to each direction of the edges there corresponds a line-shaped pattern (a string) such as the one encountered in Figure~\ref{fig:scaleinvariance}; these patterns can be computed \cite{us_solitons}. In simulations, we have observed that these strings act like the renormalization group and, thus, ensure the proportional growth of the quadratic patches in Figure~\ref{fig:scaleinvariance}. The same occurs in other sandpile models with proportional growth, which suggests that tropical geometry is  a less reductionist  tool than cellular automata to study this phenomenon.
\end{remark}

\section*{The Tropical Sandpile Model}
Here, we define a new model, the tropical sandpile (TS), reflecting structural changes when a sandpile evolves. The definition of this dynamical system is inspired by the mathematics of the previous section; TS is not a cellular automaton but it exhibits SOC.

The dynamical system lives on the convex set $\Omega = [0,N] \times [0\,N]$; we will consider $\Omega$ to be a very large square. The input data of the system is a large but finite  collection of points $P=\{p_1,\ldots,p_r\}$ with integer coordinates on the square $\Omega$. Each state of the system is an $\Omega$-tropical series (and the associated $\Omega$-tropical curve).

The initial state for the dynamical system is $F_0 = \mathbf{0}$, and its final state is the function $F_P$ defined previously. Notice that the definition of $\mathcal{F}_P$, while inspired by sandpile theory, uses no sandpiles or cellular automata whatsoever. Intermediate states $\{F_k\}_{k=1,\dots,r}$ satisfy the property that $F_k$ is not smooth at $p_1,p_2,\dots,p_k$; i.e. the corresponding tropical curve passes through these points.

In other words, the tropical curve is first attracted to the point $p_1$. Once it manages to pass through $p_1$ for the first time, it continues to try to pass through $\{p_1,p_2\}$. Once it manages to pass through $\{p_1,p_2\}$, it proceeds in the same manner towards $\{p_1,p_2,p_3\}$. This process is repeated until the curve passes through all of $P=\{p_1,\ldots, p_r\}$. 

We will call the modification $F_{k-1}\to F_{k}$ the $k$-th avalanche. It occurs as follows: To the state $F_{k-1}$ we apply the tropical operators $G_{p_1}, G_{p_2}, \ldots, G_{p_{k}}; G_{p_1}, \ldots $ in cyclic order until the function stops changing; the discreteness of the coordinates of the points in $P$ ensures that this process is finite\footnote{If the coordinates of the points in $P$ are not integers, the model is well-defined, but we need to take a limit (see \cite{us_series}), which is not suitable for computer simulations.}. Again, as before, while sandpile-inspired, the operators $G_p$ are defined entirely in terms of tropical geometry without mention of sandpiles. 

There is a dichotomy: Each application of an operator $G_p$, either does something to change the shape of the current tropical curve (in this case $G_p$ is called an active operator), or does nothing, leaving the curve intact (if $p$ already belongs to the curve). 

\begin{definition}
The size of the $k$-th avalanche is the number of distinct active operators $G_{p_i}$ used to take the system from the self-critical state $F_{k-1}$ to the next self-critical state $F_{k}$, divided by $k$. In particular, the size $s_k$ of the $k$-th avalanche is a number between zero and one, $0 \leq s_k \leq 1$, and it estimates the proportional area of the picture that changed during the avalanche.
\end{definition}

In the previous example, as the number of points in $P$ grows and becomes comparable to the number of lattice points in $\Omega$, the tropical sandpile exhibits a phase transition going into spatial SOC (fractality). This provides the first evidence in favor of SOC on the tropical sandpile model, but there is a far more subtle spatio-temporal SOC behavior that we will exhibit in the following paragraphs.

While the ordering of the points from the $1$st to the $r$-th is important for the specific details of the evolution of the system, the system's statistical behavior and final state are insensitive to it. This is called an Abelian property, which was studied extensively in \cite{MR3493110} for discrete dynamical systems (Abelian networks). Our model suggests studying  continuous dynamical systems with this Abelian property, such as, for example, Abelian networks with nodes on the plane, a continuous set of states, and an evolution rule depending on the coordinates of the nodes. We expect that the Abelian property is equivalent to the least action principle (cf. \cite{MR3493110}). 

\section*{Self-Organized Criticality in the Tropical World}

The tropical sandpile dynamics exhibit slow driving avalanching (in the sense of \cite{watkins201625} page 22).  

Once the tropical dynamical system stops after $r$ steps, we can ask ourselves what the statistical behavior of the number $N(s)$ of avalanches of size $s$ is like. We posit that the tropical dynamical system exhibits spatio-temporal SOC behavior; that is, we have a power law:
$$\log N(s) = \tau \log s+c.$$ To confirm this, we have performed experiments in the supercomputing clusters ABACUS and Xiuhcoatl at Cinvestav (Mexico City); the code is available on \cite{gitsand}. In the figure below, we see the graph of $\log N(s)$ vs. $\log s$ for the tropical (piecewise linear, continuous) sandpile dynamical system; the resulting experimental $\tau$ in this case was $\tau \sim -0.9$.

\section*{Conclusion and Further Directions}

We have obtained a piecewise-linear (continuous, tropical) model to study statistical aspects of non-linear phenomena. As tropical geometry is highly developed, it seems reasonable to believe that it will provide new reductionist explanations for physical non-linear systems in the future (as well as providing a tool for studying  proportional growth phenomena in biology and elsewhere), as has already happened with non-linear aspects of algebraic geometry and mirror symmetry.

Next, we list open questions.

From the point of view of real physical phenomena, the tropical sandpile provides a new class of mathematical and modelling tools. Some possible directions for further research are as follows:

\begin{direction}
	If $\Omega$ is a polygon with sides of rational slope, then each $\Omega$-tropical curve $C$ can be obtained as the corner locus of $F_P$ for a certain set $P$ of points. To achieve this, one considers a (finite) set $P$ of points which all belong to $C$ and cover $C$ rather densely, meaning that every point in $C$ is very close to a point in  $P$ (at a distance less than a very small $\epsilon$). As is shown in \cite{us_series}, the corner locus of $F_P$ is an $\Omega$-tropical curve passing though $P$ which solves a certain Steiner-type problem: minimizing the tropical symplectic area. However, because  the set $P$ is huge in this construction, an open question remains: Can we find a small set $P=\{p_1,\ldots,p_r \}$ (where $r$ is approximately equal to the number of faces of $C$) such that  $C$ is the corner locus of $F_P$? (We thank an anonymous referee for this question.)
\end{direction}

\begin{figure}[H]
	\centering
	\includegraphics[width=0.5\linewidth]{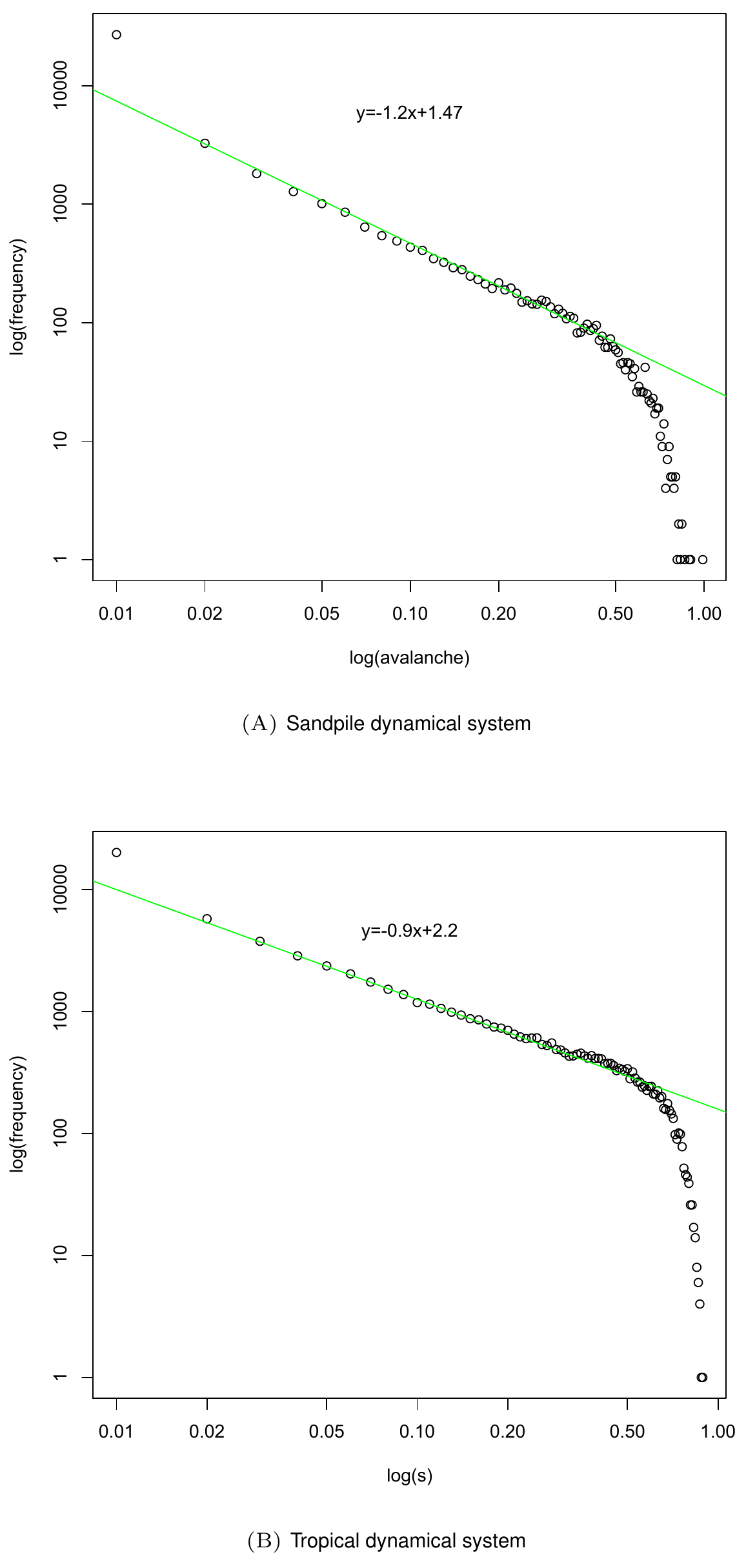}
\caption{A) The power law for sandpiles. The logarithm of the frequency  is linear with respect to the logarithm of the avalanche size, except near the right where the avalanches are larger than half of the system. Here we have $\Omega=[0,100]^2$ and is initially filled with $3$ grains everywhere, followed by $10^6$ dropped grains. B) The power-law for the tropical (piece-wise linear, continuous) dynamical system. In this computer experiment $\Omega$ has a side of $1000$ units and we add at random a set $P$ of $10000$ individual sand grains (a random large genotype).}
\end{figure}

\begin{direction}
The operators $G_p$ can be lifted to the algebraic setting but, for now, we can do this only in fields of characteristic two  \cite{us_series}. Is it possible to lift $G_p$ in characteristic zero (the complex realm)? In any case, we expect that this difficulty can be alleviated by a mirror symmetry interpretation. The issue is due to the {\it symplectic} nature of $G_p$: indeed, $\Omega$-tropical curves naturally appear as tropical symplectic degenerations \cite{mikhalkin2018examples,matessi2018lagrangian}. When $\Omega$ is not a rational polygon, one should take into account non-commutative geometry \cite{katzarkov2014definition}. What is the mirror analog of $G_p$  in the complex world? We expect that there should exist an operator $G_p^*$ acting on strings, and through this we expect power-law statistics for a mirror notion of the areas of the faces of $\Omega$-tropical curves.  

Closely related to this, in analogy to the work of Iqbar et al. \cite{MR2425292}, we conjecture that the partition function obtained by summing over statistical mechanical configurations of sandpiles should have, via mirror symmetry, an interpretation as a path integral in terms of K\"ahler geometry. Tropical geometry should play the role of toric geometry in this case. What is the precise geometric model in this situation? 

Developments in this direction would allow a new renormalization group interpretation of SOC (cf. \cite{diaz1994dynamic, ansari2008self}). 
\end{direction}

\begin{direction}
	Rastelli and K\"anel studied nanometer sized three-dimensional islands formed during epitaxial growth of semiconductors appearing as faceted pyramids that seem to be modeled by tropical series from an inspection of Figure~\ref{figpush} in this paper and Figure 3 in ~\cite{rastelli2002surface}. It would be very interesting to prove that this is so and to study the consequences of this observation to the modeling of the morphology of such phenomena.
	
	This may not be totally unrelated to allometry in biology. Is it possible that the gradient slope model, as in Section 2.5 in \cite{vollmer2017growth}, could be piecewise linear, and the corner locus slopes could prescribe the type and speed of growth for tissues?
\end{direction}

\begin{direction}
	Study the statistical distributions for the coefficients of the tropical series in Figure 6 (C) in our paper. Explain why the slopes are mostly of directions $(0,1),(1,0),(1,1),(-1,1)$. Is it possible that a concentration measure phenomenon takes place and that such a type of picture appears with probability one? (We thank Lionel Levine for this question.) 
\end{direction}

\section*{Experimental Data}

The data used to produce Figure 9 can be found in  \cite{gitsand}.

\section*{Acknowledgments}

Nikita Kalinin was funded by the SNSF PostDoc.Mobility Grant 168647, supported in part by Young Russian Mathematics award, and would like to thank Grant FORDECYT-265667 "Programa para un Avance Global e Integrado de la Matem\'atica Mexicana". Also, support from the Basic Research Program of the National Research University Higher School of Economics is gratefully acknowledged. 

Yulieth Prieto was funded by Grant FORDECYT-265667 and by ABACUS (Cinvestav). 

Mikhail Shkolnikov was supported by ISTFELLOW program.

Finally, Ernesto Lupercio would like to  thank the Moshinsky Foundation, Conacyt, FORDECYT-265667, ABACUS, Xiuhcoatl, IMATE-UNAM, Samuel Gitler International Collaboration Center and the Laboratory of Mirror Symmetry NRU HSE, RF Government grant, ag. No. 14.641.31.0001 and the kind hospitality of the University of Geneva and of the Mathematisches Forschungsinstitut Oberwolfach where this work started. \emph{In memoriam JL.}

\bibliography{bibliography}
\bibliographystyle{unsrt}
 
\end{document}